# The effect of layer number and substrate on the stability of graphene under MeV proton beam irradiation

S. Mathew [a], T.K. Chan [c], D. Zhan [d], K. Gopinadhan [a,b], A.R. Barman [a,b],

M.B.H. Breese [c], S. Dhar [a,b], Z.X. Shen [d], T. Venkatesan [a, b] and John TL Thong [a]

[a]*Department of Electrical and Computer Engineering, National University of Singapore, Singapore, 117576.*

[b]*NUSNNI-NanoCore, National University of Singapore 117576*

[c]*Center for Ion Beam Applications (CIBA), Department of Physics, National University of Singapore, Singapore 117542*

[d]*Division of Physics and Applied Physics, School of Physical and Mathematical Sciences, Nanyang Technological University, Singapore 637371*

The use of graphene electronics in space will depend on the radiation hardness of graphene. The damage threshold of graphene samples, subjected to 2 MeV $H^+$ irradiation, was found to increase with layer number and also when the graphene layer was supported by a substrate. The thermal properties of graphene as a function of the number of layers or as influenced by the substrate argue against a thermal model for the production of damage by the ion beam. We propose a model of intense electronically-stimulated surface desorption of the atoms as the most likely process for this damage mechanism.

[*] *Corresponding author:* Fax: + 65 6516 7912.
E-mail address: pmsmathew@gmail.com (S. Mathew), venky@nus.edu.sg (T. Venkatesan), elettl@nus.edu.sg ( John TL Thong).

## 1. Introduction

Graphene, a two-dimensional (2D) allotrope of carbon where carbon atoms are arranged in a hexagonal lattice, has been the subject of many fascinating studies since its isolation [1]. It is one atomic layer thick and is stable in two dimensions and even when suspended [2]. The use of graphene devices in space applications, in particular graphene based solar cells which have been already demonstrated [3], and the stability of these devices in the harsh space environment is best studied by the interaction of MeV protons with graphene. Graphene being a single atomic layer of carbon atoms is a unique system to study ion-solid interactions at the beginning of a collision cascade at the microscopic level. Understanding the mechanisms behind the defect formation and annealing in graphene is essential for defect – assisted engineering in graphene and most of the carbon allotropes [4]. The stability of $sp^2$ carbon-carbon bonds in graphene with external perturbations such as ion-irradiation as a function of layer number is required to investigate the role of dimensionality in stabilizing 2D structures like graphene. The stability of suspended graphene membranes with energetic ions is of great importance considering the recent work [5] demonstrating graphene as the ultimate membrane for ion-beam analysis of gases and other volatile systems which cannot be kept in vacuum.

Very recently, Compagini *et al.* investigated 500 keV $C^+$ irradiation effects in graphene and found that the damage in monolayer is higher than in multi layers which they attributed to a substrate effect [6]. Focused ion beams of few tens of keV have been shown to be effective in the nano-structuring of graphene. Pickard *et al.* showed that free-standing graphene ribbons of size ~5 nm can be fabricated using a 30 keV focused $He^+$ beam [7]. Lemme *et al.* also demonstrated nano-structuring and electrical isolation of graphene devices with focused keV He ion beams [8].

In this paper, we investigate the effects of graphene samples under MeV proton irradiation as a function of layer number and ion fluence for suspended and supported graphene. A mechanism is proposed for the damage mechanism arising from intense electronically stimulated desorption of the atoms.

## 2. Experimental

Graphene samples were fabricated using micro-mechanical exfoliation of Kish graphite and subsequent transfer to a SiO$_2$/Si substrate [1]. The suspended graphene samples were prepared by micromechanical exfoliation of Kish graphite onto a SiO$_2$/Si substrate with an array of pre-patterned holes prepared in the following way. Photolithography is used to transfer a mask pattern consisting of an array of holes into photo-resist spin-coated on a SiO$_2$/Si substrate. This was followed by dry etching of the exposed SiO$_2$ regions and subsequent removal of the photo-resist. The above substrate was further cleaned using oxygen plasma to remove any residual hydrocarbons remaining on the surface of the substrate. A schematic of the sample fabrication steps is shown in Fig. 1. The graphene samples prepared this way had a distribution of layer thicknesses on supported areas and also suspended regions.

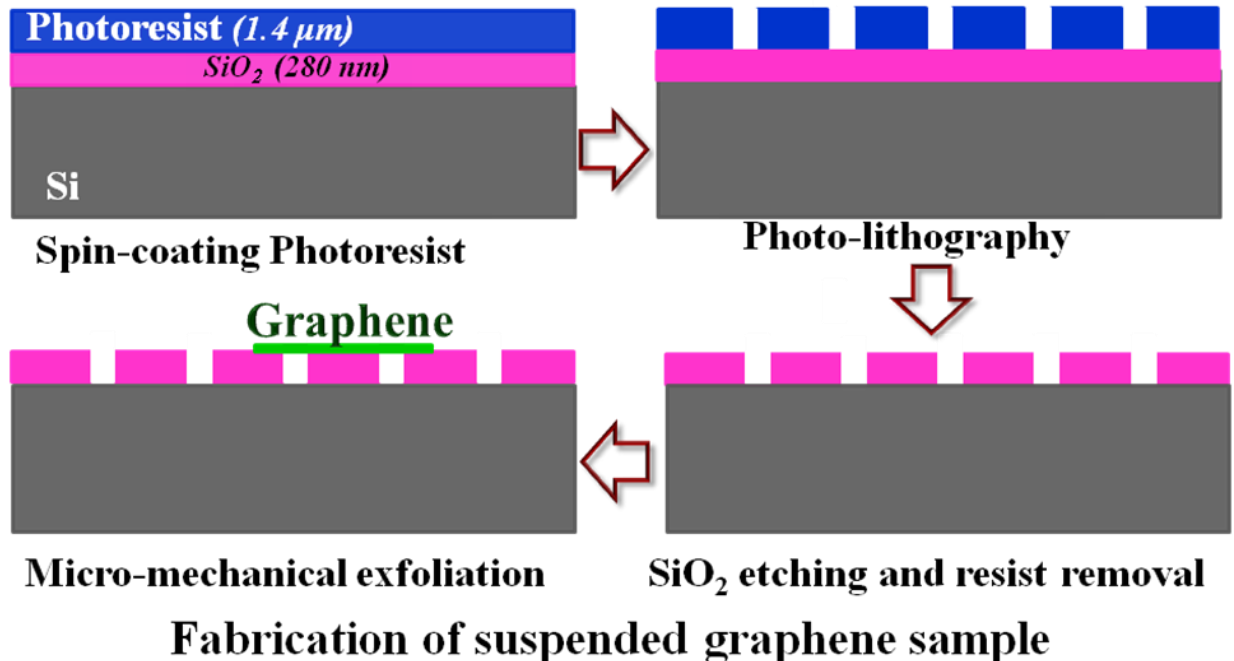

**Fabrication of suspended graphene sample**

Fig. 1. Schematic of the fabrication of suspended graphene samples.

One of the inherent technological difficulties in using exfoliated graphene samples for the ion irradiation study is the presence of contaminants and adsorbed atoms on the sample, i.e., adhesive tape residues remains on the sample (both on graphene and on $SiO_2$) and molecules from the environment adsorbing on the surface of the graphene flake. Pickard *et al.* observed ion beam induced re-deposition of hydrocarbons on graphene surface after a graphene sample with tape residues has been irradiated with focused keV He ions [7]. Formation of graphane under 5-10 keV electron irradiation of graphene with adsorbed water molecules has been reported recently [9]. Moser *et al.* probed the surface of graphene exposed to air and showed that a monolayer of water adsorbs on graphene surface and the adsorbed water does not desorb in vacuum [10]. To realize clean graphene samples for the ion irradiation study, we designed a two step annealing process: (a) annealing the exfoliated sample inside a tube furnace to remove the tape residues, and (b) heating the sample inside the irradiation chamber before each irradiation step to remove the adsorbed molecules that had been adsorbed from the ambient air. The pristine samples mentioned in the later part of the text refer to the graphene annealed using step (a) for removing the adhesive tape residues.

Ion irradiations were carried out using a 3.5 MV Singletron facility at the Center for Ion Beam Applications at the National University of Singapore. The graphene samples were loaded into the nuclear microscopy chamber with a strip heater attached in the sample holder for the *in situ* heating procedure mentioned earlier. A collimated beam of 2 MeV protons was focused to a beam spot size of ~ 5 μm on target using a set of quadrupole lenses. An optical microscope attached to the irradiation chamber was used to locate the graphene flake in the sample. The focused ion beam was then raster-scanned under normal incidence over an area of $800 \times 800$ $\mu m^2$ with the irradiated graphene flake positioned at the centre of each scan. The pressure in the irradiation chamber during the irradiation was $1 \times 10^{-6}$ mbar. The ion beam current density was kept at 0.5 pA/$\mu m^2$ for ion fluences up to $1 \times 10^{18}$ ions/cm$^2$, and 1.3 pA/$\mu m^2$ for ion fluences $6 \times 10^{18}$ ions/cm$^2$ and above. Very recently, Ramaos $et\ al.$ measured the surface temperature of highly oriented pyrolytic graphite (HOPG) under various ions of energy 2-25 MeV using thermal imaging of HOPG during the ion irradiation [11]. The estimated temperature is in the range of 100-140 °C. In our system, the 2 MeV protons dissipate most of its energy in the substrate silicon. Assuming an adiabatic system, and using the beam parameters involved in present study, the estimated temperature in our sample is around 100 °C. Hence the possibility of annealing of the defects during the irradiation has not been considered significant in the present study. Visible Raman spectroscopy (excitation radiation 532 nm) and imaging were carried out using a WITec CRM200 Raman system. The Raman spectrum is analyzed by curve fitting using multiple Lorentzians with a sloping background. Atomic Force Microscope (AFM) imaging of the irradiated samples was carried out in the tapping mode using an Agilent 5500 AFM system.

## 3. Results and Discussion

Irradiations with 2 MeV protons at different fluences were carried out on a graphene sample with a region encompassing 1, 2, and 4 layers. An optical micrograph of the sample is shown in Fig. 2(a). The layer thickness and uniformity were confirmed by using Raman imaging of the flake. The differences in layer number are clear from the Raman image based on the full width at half maximum (FWHM) of the 2D band shown in Fig. 2 (b) as discussed below.

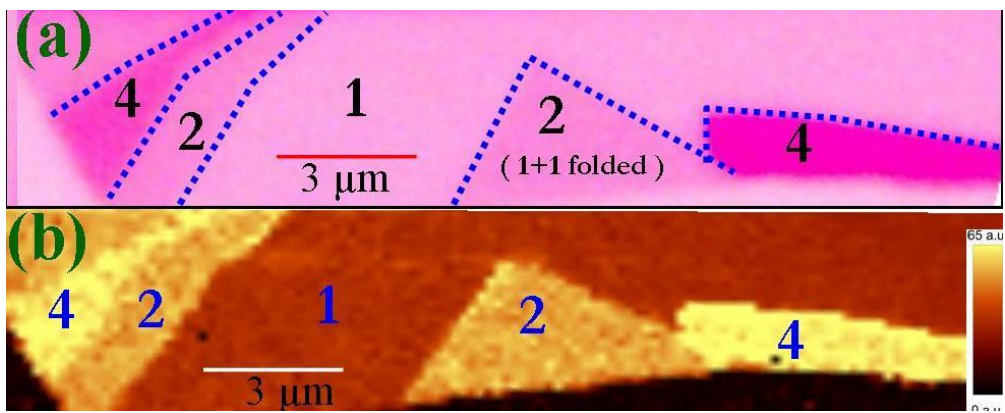

Fig. 2-(a) Optical micrograph of the graphene flake with 1, 2, and 4 layer graphene layer (b) the corresponding Raman microscopy image using the FWHM of 2D peak. The graphene layer numbers are labeled in both (a) and (b).

Raman spectrum of a pristine monolayer graphene flake is shown in Fig. 3A(a). The prominent Raman modes in Fig. 3A(a) are the G mode at 1587 cm$^{-1}$ and the 2D mode at 2669 cm$^{-1}$ respectively [12]. The FWHM of the 2D peak is 33 cm$^{-1}$ which corresponds to a monolayer graphene [13]. For an undoped graphene sample, the intensity of the 2D peak will be about 4 times that of G peak [12]. A reduction in the intensity of 2D peak in annealed and air exposed graphene samples is a common feature due to the intrinsic hole doping effect from the ambient air, as reported by Ni *et al* [14].

Monolayer graphene irradiated at a threshold fluence of $\sim 1 \times 10^{16}$ ions/cm$^2$ begins to show a D mode at 1350 cm$^{-1}$ (result not shown here). This is the in-plane breathing mode of $A_{1g}$ symmetry due to the presence of six-fold aromatic rings and requires a defect for its activation [12]. The ratio of the integrated intensities of D to that of G (denoted as I(D)/I(G)) increases with fluence. On the irradiated samples, apart from D, G, and 2D modes, another peak at 2930 cm$^{-1}$ which is a combination mode of D and G is also visible [12]. As the fluence increases, the second order peaks increase in width and in Fig. 3A(d) those peaks are barely seen. The deconvolution of the spectrum in the irradiated samples shows a sharp mode at 1623 cm$^{-1}$ called the D' mode [12] and extra broad features at 1460 and 1555 cm$^{-1}$ (hereafter these modes are denoted as $f_1$ and $f_2$ modes respectively).

The Raman spectra of pristine and irradiated, bi-layer, 4 layers and graphite are shown in Figs. 3 B, C and D respectively. The I(D)/I(G) ratio is found to increase with ion fluence in all of the irradiated samples (panels (b)- (d) in Fig. 3 B, C and D). In Fig. 3C(d) the D peak is found to be highly asymmetric, and a broad band at 1311 cm$^{-1}$ is clearly visible from the fitted data in the inset. The $f_1$ and $f_2$ peaks start to appear only at a fluence of $1 \times 10^{18}$ ions/cm$^2$ in both bilayer and 4-layer samples; in graphite these modes are present only in Fig. 3D(d) at a fluence of $6 \times 10^{18}$ ions/cm$^2$. The D' mode is present in few-layer graphene and graphite samples irradiated at a fluence of $1 \times 10^{18}$ ions/cm$^2$ and above.

The fluence dependence of the damage as measured by the Raman spectra of pristine and irradiated monolayer and multilayer graphene is shown in Fig. 3E. The variation of I(D)/I(G) ratio with ion fluence $\varphi$ can be fitted using the following function f($\varphi$)

$$f(\varphi) = \alpha[1 - e^{-(\varphi/\varphi_0)}] \qquad (1)$$

where $\alpha$ and $\varphi_0$ are two fitting parameters. The best fitted curves with experimental data are shown in Fig. 3E. The parameter $\alpha$ being a fixed value, the non-linearity in defect production comes from the second factor in Eq. (1), which essentially presents the probability of generating a defect at a given ion fluence. The parameter $(\varphi_0)^{-1}$ represents damage cross section for the impactof a single ion. From Fig. 3E, it can be seen that the value of $(\varphi_0)^{-1}$ for a monolayer is one order higher than that of few-layer graphene samples. On graphite, the curve shows a saturation behavior, which can be due to the following reason. The graphene layers down to 50 nm contribute to the intensity of the G peak in our present Raman study. The D peak originates from the damaged layers and most of the damage is concentrated at the surface within a few layers.

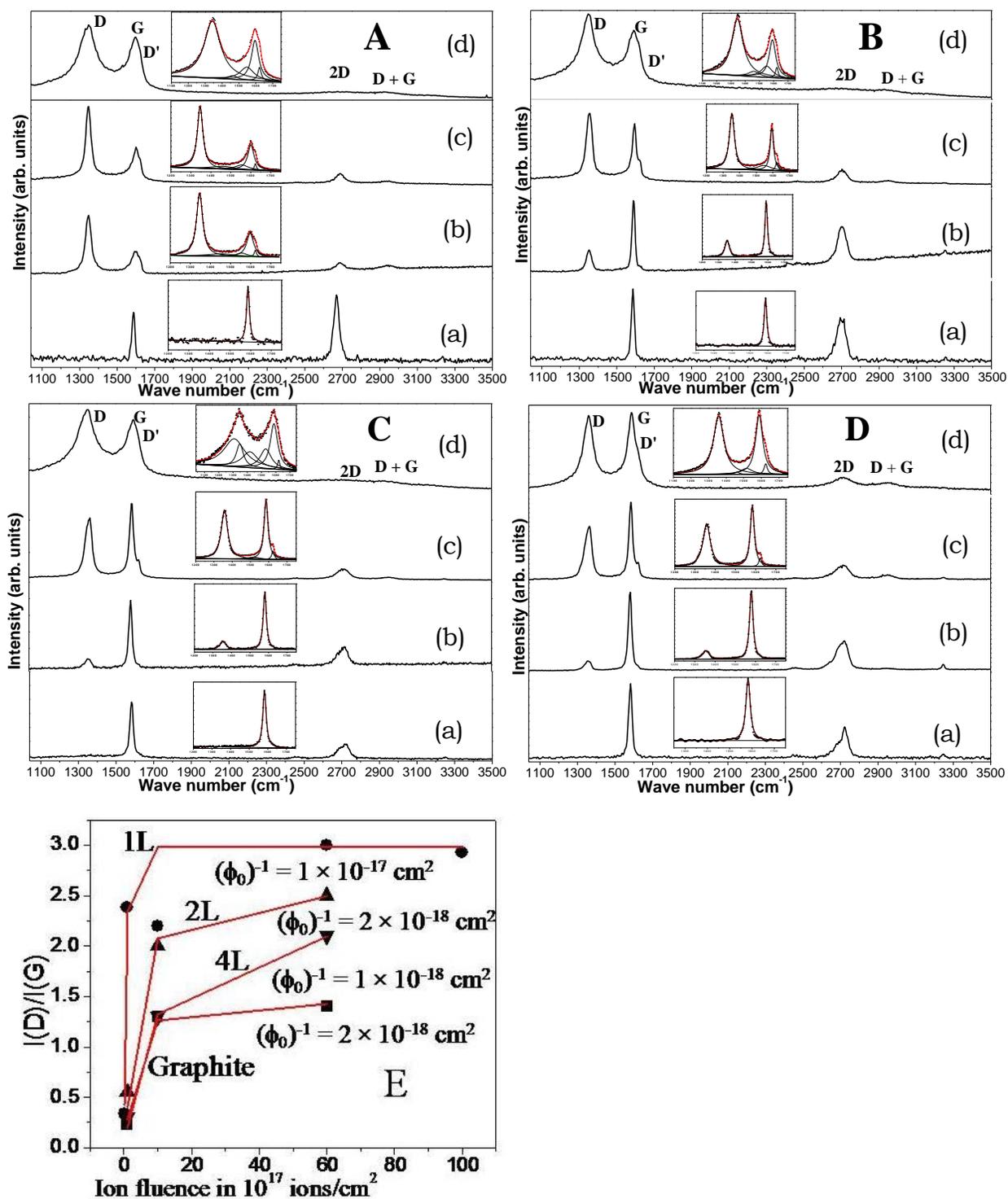

Fig. 3- Panel A-Raman spectrum from (a) pristine monolayer graphene and the same sample irradiated at fluences of (b) $1 \times 10^{17}$ ions/cm$^2$, (c) $1 \times 10^{18}$ ions/cm$^2$ and (d) $6 \times 10^{18}$ ions/cm$^2$, panels B, C and D correspond to the same for a 2 layer graphene, 4 layer graphene and Graphite

respectively. The inset shows the fitted spectrum with experimental points. The variation of I(D)/I(G) for a monolayer, 2 layer, 4 layer and graphite sample with ion fluence is shown in panel E. In panel E, the spectra is fitted with the following function $f(\varphi) = \alpha[1-e^{-(\varphi/\varphi_0)}]$.

The most probable defects expected by ion irradiation in few-layer graphene and graphite are vacancies and interstitials since the threshold energy required to produce in-plane knock-on collision needs higher transferred energies than for the off-plane displacements in highly anisotropic carbon materials like graphite [15,4]. Very recently Lehtinen *et al.* showed the importance of in-plane recoils for the amorphization of monolayer graphene and 2D materials [5]. It has been observed that single-walled carbon nanotube (SWCNT) is more sensitive to charged-particle irradiation than multi-walled carbon nanotubes (MWCNTs) [16,4]. The threshold for atom displacement for a SWCNT is reported to be lower than that of MWCNT [16]. The bridging of graphene planes in graphite and MWCNT due to the defects produced by ion or electron irradiation have been demonstrated to be the energetically favorable configuration [15-17,4]. The formation of inter-layer covalent bonding has been found to be the most appropriate mechanism for bridging inter-layer graphene layers in graphite and MWCNT [4,16,18-20]. A gauge for the presence of covalent bonds formed in the irradiated samples can be obtained using UV-Raman spectroscopy. Visible Raman spectroscopy is 50-230 times more sensitive to $sp^2$ sites compared to $sp^3$ sites as visible photons preferentially excite the $\pi$-states (exciting $\sigma$ states of the $sp^3$ sites requires higher photon energy) [21]. The UV Raman spectroscopy results (results not shown here) do not show the presence of diamond-like bond in the irradiated samples.

Let us now explore the origin of the damage in the irradiation process. The thermal conductivity ($\kappa$) [22] and heat capacity [23] of graphene both increase as the number of layers of graphene decrease. Hence this does not favor a thermal model for the damage since the damage threshold would otherwise decrease with increasing thickness. The sputtering of unsaturated carbon atoms surrounding vacancies on monolayer graphene due to the ballistic transfer of energy from 80 kV electron beam has been reported [24]. An idea about the contribution of ballistic effects in the present MeV proton irradiation on graphene system can be estimated by calculating the displacements per target atoms (dpa) using TRIM simulations. If we consider sputtering due to head-on collisions, the calculated displacements per atom from TRIM simulations [25] yield about 0.04 dpa at a fluence of $10^{19}$ cm$^{-2}$, which is an order of magnitude too small compared to the dose at which we see a full disruption of a graphene layer (Fig. 5, below). For TRIM simulations, the sample is treated as an amorphous matrix with homogenous mass density and the ion kinetic energy is transferred ballistically to the target atom. Very recently Lehtinen *et al.* showed the importance of incorporating the actual atomic structure for estimating the irradiation damage in 2D systems such as graphene [5]. Also TRIM simulation treats the dissipation of transferred energy in a 3D system, whereas for the case of a 2D system like graphene the transferred energy is dissipated in a two-dimensional plane. Krasheninnikov *et al.* investigated the microscopic mechanism of collisions between energetic protons and graphene nano-structures and showed that the electronic and ionic degree of freedom accommodates the transferred energy as a function of impact parameter and projectile energy [26]. The energy loss in electronic stopping is 32 eV/nm in graphite and it is quite possible that at this energy atoms are stimulated to desorb from the surface. This then leaves electronically-stimulated surface desorption as the most likely mechanism. If this were the case, then we would expect a significant difference between the damage threshold for a graphene layer which is free standing versus one that is supported on a substrate. In the framework of the thermal spike model, one can expect a reduction in the induced defects in suspended graphene compared to a supported one due to the reduced thermal conductivity of supported graphene [27]. We have fabricated a sample with one and three layer graphene suspended over the pre-patterned holes in $SiO_2$ and compared the induced defects in both suspended and supported graphene. On the supported graphene samples discussed above, the amorphous $SiO_2$ present at the interface has a vital role in generating the collision cascade at the interface (mainly due to the recoiled carbon atoms from graphene). The influence of the backscattered protons from the substrate silicon (cross-section for this processes is ~0.06 barns per steradian) can be safely ignored compared to the impact of $SiO_2$ at the interface for inducing damage in supported graphene samples.

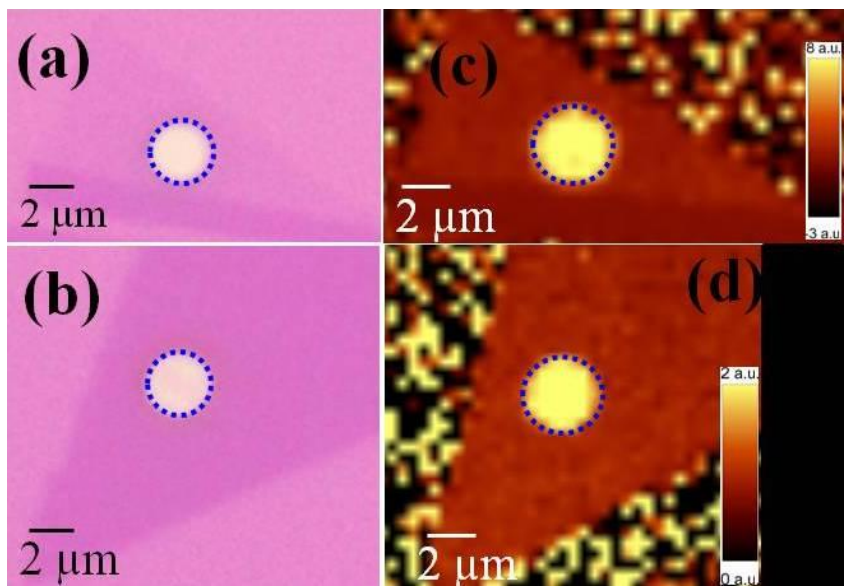

Fig. 4- Optical micrograph of suspended (a) monolayer graphene sample (b) three layer graphene sample. The corresponding Raman Microscopy image created using the I(2D)/I(G) ratio of (c)

monolayer, (d) 3-layer graphene sample. The suspended graphene region is marked using a dashed circle in all the panels.

An optical micrograph of the suspended graphene samples is shown in Fig. 4(a) and (b). The suspended graphene regions are marked using dashed circles. Raman microscopy images showing the I(2D)/I(G) ratio of monolayer and 3 layer regions are given in Figs. 4 (c) and (d). The intense signal (from the colour code) in Figs. 4(c) and (d) at the suspended region shows that the graphene remain free-standing over the etched hole in $SiO_2$ [28].

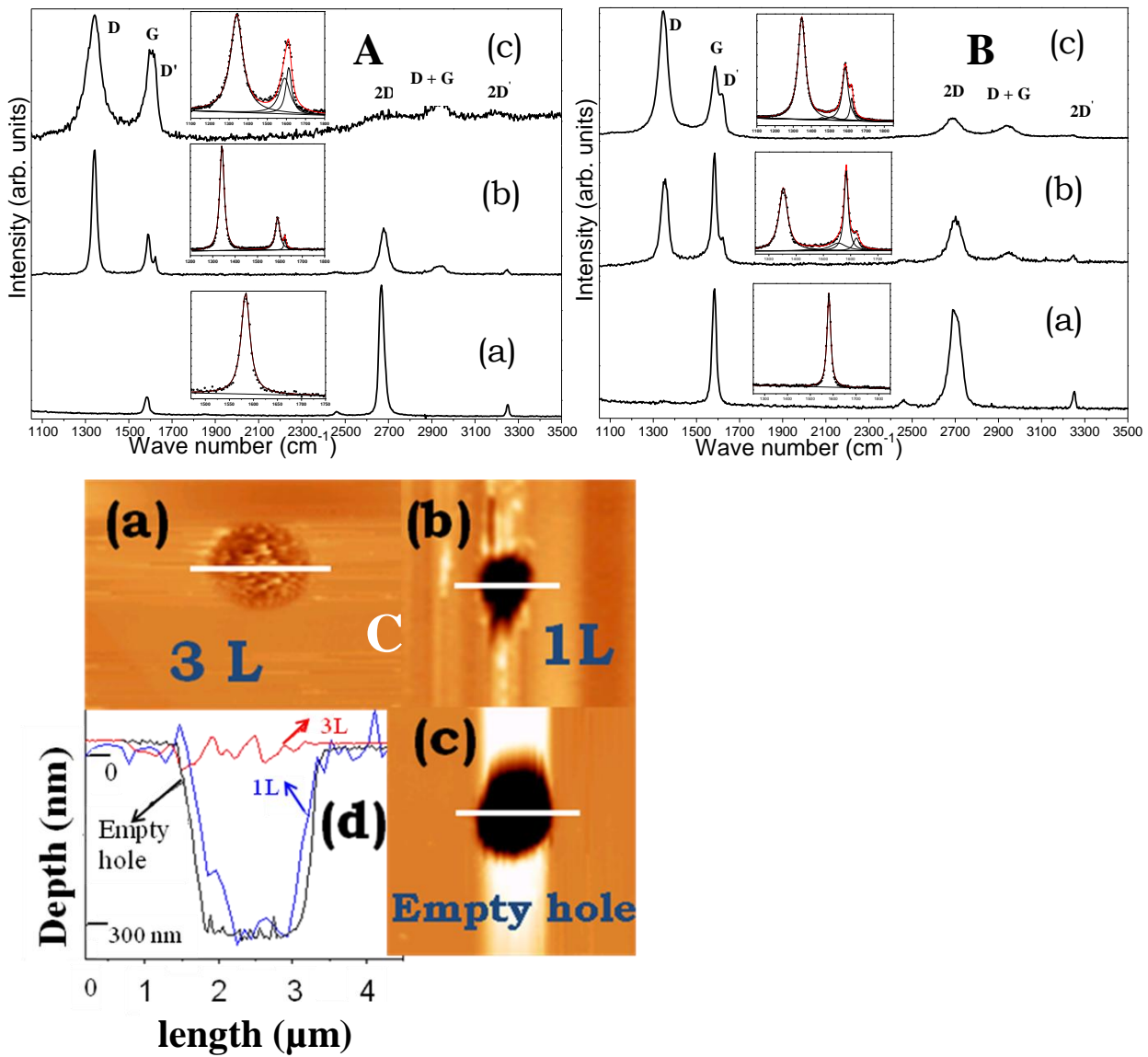

Fig. 5- Panel A - Raman spectrum from (a) a pristine monolayer suspended graphene and the same sample irradiated at fluences of (b) $1 \times 10^{18}$ ions/cm$^2$ and (c) $1 \times 10^{19}$ ions/cm$^2$, panel B corresponds to the the same for 3 layer graphene sample. The inset shows the fitted spectrum with experimental points. Panel C: Atomic force microscope image of (a) monolayer suspended graphene sample and (b) 3 layer suspended graphene sample irradiated at a fluence of $1 \times 10^{19}$ ions/cm$^2$ and (c) an empty etched hole in SiO$_2$ substrate. The corresponding line profiles (red-3 layer, blue- 1 layer and black for an empty hole on SiO$_2$) are shown in (d).

The Raman spectra of the pristine and irradiated suspended monolayer and 3-layer samples are shown in Figs. 5A and 5B respectively. The FWHM of the 2D peak in the pristine samples are 26, 59 cm$^{-1}$ respectively, which corresponds to mono and ~3 layer graphene [13]. The I(D)/I(G) ratio and thus the induced defects in the suspended monolayer graphene under low fluence irradiation is much higher than that of the supported region. The variation of the I(D)/I(G) ratio with ion fluence and layer number is shown in the Table 1. AFM on the sample irradiated at a fluence of $1 \times 10^{19}$ ions/cm$^2$ shows that the 3 layer graphene remains suspended while the monolayer has collapsed into the etched hole from the line profiles shown in Fig. 5C(d). It can be clearly seen that $f_1$ and $f_2$ peaks are absent on the irradiated suspended monolayer sample (The origins of the $f_1$ and $f_2$ modes will be discussed elsewhere). Also we had shown in Fig. 3 that the induced defects in monolayer are more than those of few–layer graphene at all of the fluences.

Table 1. I(D)/I(G) ratio at various ion fluence of suspended and corresponding supported graphene flake for a mono layer and 3 layer sample.

| Fluence (ions/cm$^2$) | 1 layer | | 3 layer | |
|---|---|---|---|---|
| | Suspended | Supported | Suspended | Supported |
| $1 \times 10^{18}$ | 3.3 | 2.1 | 1.3 | 1.4 |
| $1 \times 10^{19}$ | 3.1 | 3.0 | 2.4 | 2.5 |

These results in conjunction with the observation of higher I(D)/I(G) ratio in the suspended monolayer compared to supported region clearly demonstrate that the graphene-graphene interaction along the third dimension makes the quasi-two-dimensional graphene more stable. If thermal effects dominate the defect creation processes, one would expect a higher amount of damage in the few layers compared to that of a monolayer. Also the defects in supported graphene would have been considerably higher than those of suspended graphene (κ of suspended graphene is 5 times that of supported graphene [27]). Considering these aspects we can safely ignore the thermal spike model.

The velocity of ions used in this study is one order higher than that of the Fermi velocity of electrons in graphene ($1 \times 10^6$ m/s). The electronic excitations in graphene play a major role in this energy range [26,29,30]. The decay of the electronic excitations in graphene is different from that of conventional free-electron-like metals even though both are electrically conducting systems. The reduced electron density makes the screening of *e-e* interactions in graphene much smaller than that of metals [31]. Lenner *et al.* demonstrated the breakage of the in-plane carbon bonds in graphene due to excitations of the $\pi$-electrons [32]. The disruption of graphene lattice due to non-thermal decay of the excited electrons has also been reported very recently [33]. The increase in the energy of electronic degree of freedom leads to modified inter-atomic forces and subsequent motion of the target atoms leading to electronically stimulated desorption. Thus the lattice relaxation/electronically stimulated desorption model can be more appropriate in describing the MeV proton induced defects in graphene [26,29-34].

Thus the fact that the threshold is lower for the suspended graphene as opposed to the supported graphene supports our hypothesis of surface desorption under intense electronic excitation. This also explains why multilayer graphene has a higher threshold since the surface to volume ratio is reduced for the thicker layers. Recently ferromagnetic (or ferri-magnetic) ordering in highly oriented pyrolitic graphite was seen under MeV proton irradiation [11] and 80% of the measured magnetic signal was found to originate from the first 10 nm of the surface [35]. This observation indicates that the defects induced by MeV protons in mono and few-layer graphene tend to be localized at the surface which also supports our model of electronically stimulated desorption.

## 4. Conclusions

The stability of graphene was found to grow with increase in layer number and this points towards the role of interaction along the third dimension in stabilizing the quasi two-dimensional graphene. The analysis of the evolution of the defects with ion fluence has shown that the damage cross-section for monolayer is one order higher than that for few layers. The electronically stimulated surface desorption of the atoms (the lattice-relaxation model) is found to be appropriate for explaining the nature of the ion induced damage in graphene under MeV proton irradiation. This model is consistent with the lower damage threshold for suspended graphene with respect to supported graphene and also the reduced damage at a given dose with increasing layer number.